\documentstyle[aps,psfig,twocolumn]{revtex}
\begin{document}
\draft \flushbottom
\twocolumn[
\hsize\textwidth\columnwidth\hsize\csname
@twocolumnfalse\endcsname
\title{Nernst Effect in Electron-Doped Pr$_{2-x}$Ce$_{x}$CuO$_4$}
\author{Hamza Balci, C.P.Hill, M. M. Qazilbash, and R.~L.~Greene}
\address{1. Center for Superconductivity Research,
Department of Physics, University of Maryland,
College~Park, MD-20742}
\date{\today}
\maketitle
\tightenlines
\widetext
\advance\leftskip by 57pt
\advance\rightskip by 57pt

\begin{abstract}
The Nernst effect of Pr$_{2-x}$Ce$_{x}$CuO$_4$ (x=0.13, 0.15, and
0.17) has been measured on thin film samples between 5-120 K and
0-14 T. In comparison to recent measurements on hole-doped
cuprates that showed an anomalously large Nernst effect above the
resistive T$_c$ and H$_{c2}$~\cite{xu,wang1,wang2,capan}, we find
a normal Nernst effect above T$_c$ and H$_{c2}$ for all dopings.
The lack of an anomalous Nernst effect in the electron-doped
compounds supports the models that explain this effect in terms of
amplitude and phase fluctuations in the hole-doped cuprates. In
addition, the H$_{c2}$(T) determined from the Nernst effect shows
a conventional behavior for all dopings. The energy gap determined
from H$_{c2}$(0) decreases as the system goes from under-doping to
over-dopingin agreement with the recent tunnelling experiments.
\end{abstract}
\pacs{} ] \narrowtext \tightenlines {\bf INTRODUCTION}
\newline
Recent Nernst effect measurements~\cite{xu,wang1,wang2,capan} on
hole-doped cuprate high-T$_c$ superconductors have shown very
surprising results. Especially in the under-doped regime of these
cuprates, an anomalous Nernst signal has been observed to persist
to temperatures up to 50-100 K above T$_c$, and to magnetic fields
much larger than the resistive H$_{c2}$.  The authors have
interpreted this anomalous signal above the conventional T$_c$ or
H$_{c2}$ (the T$_c$ or H$_{c2}$ of resistivity and magnetization)
as evidence for vortex-like excitations, and have defined a new
T$_c$ and H$_{c2}$ for Cooper pair formation in these compounds.
In this picture there is a temperature (or field) at which the
Cooper pairs start to form, and another temperature(or field)
below which the Cooper pairs attain phase coherence throughout the
sample. Therefore, the T$_c$(or H$_{c2}$) of resistivity
measurements corresponds to the temperature (or field) that
coherence has been obtained, whereas the onset of the anomalous
Nernst signal corresponds to the temperature (or field) of the
Cooper pair formation.  The authors also suggest that this
anomalous Nernst signal is related to the pseudogap or some
interaction between the pseudogap state and the superconducting
state since the Nernst signal follows a pattern similar to the
pseudogap phase diagram(i.e. the signal above T$_c$ is more
pronounced in the under-doped regime)~\cite{timusk}.

These Nernst effect measurements have inspired a revisit to the
theory of superconducting fluctuations in the cuprates. These
theoretical studies have proposed that the anomalous Nernst effect
can be explained in terms of various types of fluctuations or in
terms of a preformed pair model. Kontani suggests that including
antiferromagnetic fluctuations in addition to superconducting
fluctuations in the under-doped regime would explain the unusually
large Nernst signal above T$_c$~\cite{kontani}. Ussishkin \emph{et
al.} ~\cite{ussishkin} suggest that Gaussian superconducting
fluctuations above T$_c$ are able to explain the Nernst effect for
the optimally-doped and over-doped regimes. For the under-doped
regime they suggest that strong non-Gaussian fluctuations reduce
the mean-field transition temperature T$^{MF}_c$ and therefore the
mean field T$^{MF}_c$ should be used in calculations instead of
the actual T$_c$ in order to take into account the contribution of
the non-Gaussian fluctuations to the Nernst
effect~\cite{ussishkin}. Another proposal came from Tan \emph{et
al.} ~\cite{tan} in which they proposed a preformed pair
alternative to the vortex-like excitations scenario to explain the
anomalous Nernst effect in the under-doped hole-doped cuprates.
The work of Carlson \emph{et al.}~\cite{carlson} is another
important study about the nature of superconducting fluctuations
that we should mention.  In this study the cuprates are classified
in terms of their pairing strength (a measure of the
superconducting gap) and phase stiffness (a measure of the
superfluid density). This theory would predict that in the
hole-doped cuprates the fluctuations in the phase of the order
parameter would dominate the Nernst signal up to a certain
temperature above T$_c$, and at still higher temperatures there
should be contributions to the Nernst effect from fluctuations
both in phase and the amplitude of the order parameter (Gaussian
fluctuations). The same study would predict that these
fluctuations should be much less in the electron-doped cuprates.
At present, none of the proposed explanations for the large Nernst
signal observed in the hole-doped compounds have gained general
acceptance.

Early measurements on hole-doped cuprates, which were concentrated
on the optimally-doped regime, showed a large Nernst signal below
T$_c$ (the well known vortex Nernst effect) which diminished
rapidly close to T$_c$ (H$_{c2}$), and merged to the normal state
Nernst signal~\cite{zeh}. This behavior was similar to that
observed in conventional superconductors, except for a broader
fluctuation regime. The Nernst effect studies in the
electron-doped cuprate superconductors(all previous measurements
were on Nd$_{1.85}$Ce$_{0.15}$CuO$_{4-\delta}$(NCCO)) showed the
same behavior in the superconducting state. However, the normal
state behavior was quite different~\cite{jiang,fournier,gollnik}.
An anomalously large Nernst voltage in the normal state was
interpreted as evidence for the existence of two types of
carriers, not vortex-like excitations. The two carrier
interpretation has recently been supported for optimal doping by
ARPES measurements which showed electron pockets on a hole-like
Fermi surface~\cite{armitage}. The doping dependence of the Nernst
effect in the electron doped superconductors was studied by
varying the oxygen content of NCCO, but the cerium doping
dependence was not investigated.

In this paper we report Nernst effect data for the electron-doped
superconductor Pr$_{2-x}$Ce$_{x}$CuO$_4$(PCCO) at different cerium
dopings, and discuss some of the important issues that were raised
by the recent Nernst effect measurements on the hole-doped
compounds. Magnetic field and temperature dependence of the Nernst
voltage, and temperature dependence of H$_{c2}$ close to T$_{c}$
are presented. In addition, H$_{c2}$ values obtained from Nernst
effect and resistivity are compared. Unlike the recent results on
some hole-doped compounds~\cite{xu,wang1,wang2,capan}, our data
does not show an anomalous Nernst signal above T$_c$(or H$_{c2}$)
for the optimally-doped and over-doped compounds. However, the
under-doped compound shows a larger fluctuation regime. The
H$_{c2}$(T) obtained from the Nernst effect follows a conventional
linear temperature dependence close to T$_c$ for all dopings we
studied in contrast to an anomalous curvature found in many
previous resistivity determinations of H$_{c2}$(T). The critical
field, H$_{c2}$(0), and the superconducting energy gap deduced
from H$_{c2}$(0) increase with decreasing doping even though T$_c$
has a different doping dependence. The magnitude of the Nernst
signal in the normal state is very similar for different cerium
dopings. It is too large to be explained by a one carrier
(one-band) model and it does not show the temperature dependence
to be caused by vortex-like excitations or superconducting
fluctuations. This suggests that two types of carriers (bands)
exist in all the cerium dopings we studied and they are the origin
of the large Nernst signal above T$_c$.
\newline
\newline

{\bf THEORETICAL BACKGROUND}

The Nernst effect is a thermomagnetic effect, in which a
transverse potential difference is induced in the presence of a
longitudinal thermal gradient and a perpendicular magnetic field.
Since a detailed account of the theory of Nernst effect is given
in literature ~\cite{wang1,gollnik,clayhold}, we will not repeat
it in this manuscript except for a brief summary. The Nernst
effect of a normal metal with one band of conduction is known to
be small (due to Sondheimer's cancellation ~\cite{sondheimer}),
and this effect is linear in magnetic field since it is induced by
the Lorentz force on the charge carriers ($F=q v \times B$). On
the other hand, the Nernst effect in the mixed state of a
superconductor is due to vortex motion (rather than electrons or
holes of a normal metal). A vortex moving in a magnetic field
induces an electric field transverse to its motion due to aphase
slip effect ~\cite{josephson}. The vortex Nernst effect is the
dominant thermomagnetic effect and it is much larger than the
normal state Nernst effect. In a conventional type-II
superconductor the vortex Nernst effect goes through a peak when
the magnetic field is scanned at a constant temperature and it
merges onto the small normal state Nernst signal around H$_{c2}$.
The Nernst effect will be represented as $e_y = E_y / \nabla T$ in
the rest of the manuscript.

There are several issues worth mentioning that are special to the
normal state of electron-doped compounds. As we said earlier, the
Nernst effect of a normal metal with one band of conduction is
small. However, the Nernst effect in the normal state of the
electron-doped cuprates was observed to be large ($\sim$ two
orders of magnitude larger than expected for a one band system).
This observation combined with unusually small magnitude for the
ratio of Hall angle to the thermal Hall angle (again roughly two
orders of magnitude smaller than expected for a one band system)
were interpreted as evidences for the existence of two-bands of
conduction in these materials ~\cite{jiang,fournier,gollnik}.
\newline
\newline
{\bf SAMPLES AND EXPERIMENTAL SETUP}
\newline
The measurements
were performed on Pr$_{2-x}$Ce$_{x}$CuO$_4$ (x=0.13, 0.15, and
0.17) thin films grown by the pulsed laser deposition technique on
STO substrates.  The thickness of the films was between 2000-3000
A. The sample was attached on one end to a copper block with a
mechanical clamp (for better thermal contact), with the other end
left free (similar to a diving board-see Ref.~\cite{fournier} for
a figure). A temperature gradient was created by heating up the
free end with a small heater attached on the film. Two Lakeshore
cernox thermometers were attached on the two ends of the sample to
monitor the temperature gradient continuously. The temperature
gradient was between 1-2.5 K/cm depending on the temperature of
the measurement. The temperature of the sample was determined by
taking the average of the temperatures at the hot and cold sides.
The measurements were performed under vacuum, and the magnetic
field was perpendicular to the ab-plane. The Nernst voltage was
measured with a Keithley 2182 Nanovoltmeter which has a
sensitivity of several nanovolts. The measurements were made at
fixed temperatures while the field is scanned slowly at a rate of
~20 Oe/sec. The temperature stability was a few millikelvins
during the field scan. The Nernst signal is measured at positive
and negative field polarity, and (1/2) the difference of the two
polarities is taken to remove any thermopower contribution.
\newline
\newline
{\bf DATA AND ANALYSIS} \newline Fig.1 shows the resistivity data
for the films used in this study. The T$_c$, the sharpness of the
superconducting transition, and the behavior of resistivity in
high magnetic fields below the zero-field T$_c$(insulating-like
for optimally-doped and under-doped samples, and metallic for
over-doped samples) show the high quality of the
films~\cite{fournier}.

Fig.2-a, -b, -c show the low temperature Nernst effect data for
the three dopings we studied, and Fig.2-d shows a comparison of
the Nernst signal for the three dopings at T/T$_c\approx$ 0.7. The
superconducting vortex Nernst signal of the over-doped and
optimally-doped samples crosses over to the normal-state Nernst
signal(linear in magnetic field) very close to the resistive
H$_{c2}$ . In the under-doped sample, the transition from the
superconducting state to normal state occurs over a wider field
range suggesting that the fluctuation regime is broader for the
under-doped regime compared to the other dopings.

\begin{figure}
\centerline{\psfig{figure=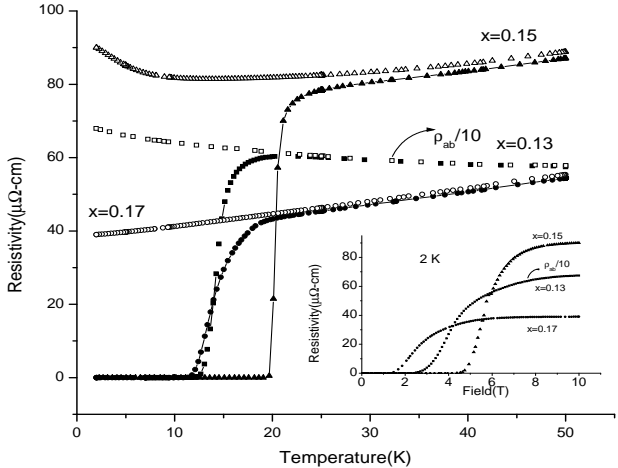,width=9.0cm,height=7.0cm,clip=}}
\caption{Resistivity of the optimal, over, and under-doped PCCO as
a function of temperature at zero field (dark symbols) and H=14 T
(open symbols). The inset shows the resistivity of the same
samples as a function of field at T=2 K.}
\end{figure}

\begin{figure}
\centerline{\psfig{figure=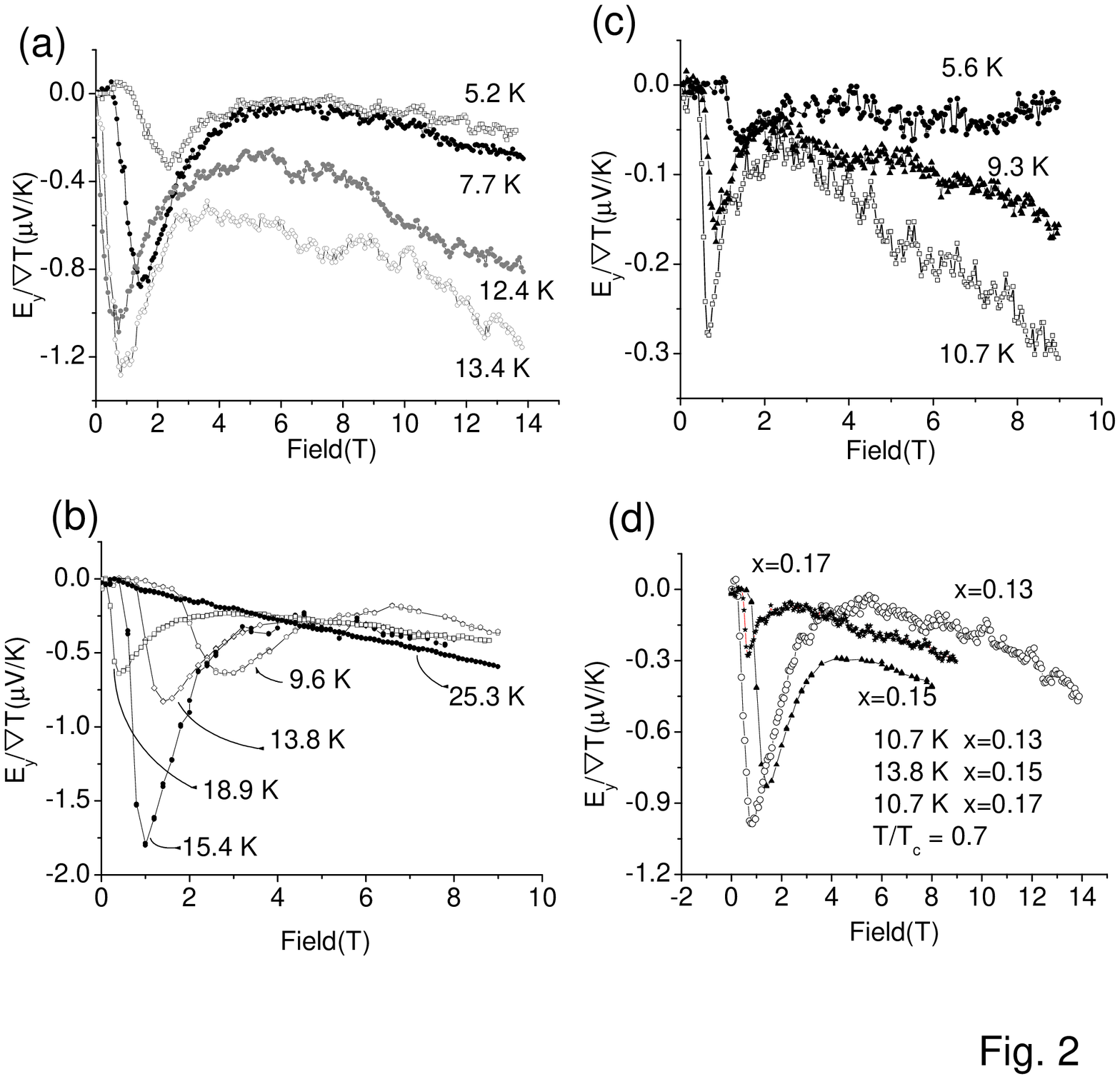,width=10.0cm,height=8.0cm,clip=}}
\caption{The low temperature Nernst effect of \textbf{(a)} the
under-doped, \textbf{(b)} optimally-doped, \textbf{(c)} over-doped
samples as a function of magnetic field at fixed temperatures.
\textbf{(d)} shows a comparison of the three dopings at roughly
the same T/T$_c$=0.7.}
\end{figure}

 Nevertheless in
all three dopings the Nernst signal behaves very differently from
the hole-doped cuprates in which an anomalous Nernst signal has
been observed ~\cite{xu,wang1,wang2,capan}. In the electron-doped
PCCO the peak of the vortex Nernst signal is quite sharp in all
the dopings we studied. However, these hole-doped
compounds~\cite{xu,wang1,wang2,capan}, particularly the
under-doped compounds, show an extended peak for the vortex Nernst
signal that persists to fields much larger than the resistive
H$_{c2}$ even at temperatures very close to $T_c$.

Fig.3-a shows the typical Nernst effect for T$>$T$_c$ for the
optimally-doped sample. The linear field dependence of the charge
carrier Nernst effect is clearly seen, and no anomalous behavior
is observed even at temperatures very close to the resistive
T$_c$. Under-doped and over-doped samples behave very similarly to
the optimally-doped sample, therefore the data for these dopings
is not shown here. Fig.3-b summarizes the temperature dependence
of the Nernst signal at 9 T for T$>$T$_c$. The dome-like behavior
that was observed in Nd$_{1.85}$Ce$_{0.15}$CuO$_{4-\delta}$ for
different oxygen dopings~\cite{fournier,gollnik} is also observed
in PCCO for different cerium dopings. The large magnitude of the
Nernst signal is also similar to that observed in NCCO for
T$>$T$_c$. This large magnitude of the Nernst signal and some
other observations that are discussed in detail in
Ref.~\cite{fournier} were interpreted as evidence for the
existence of two-types of carriers in electron-doped cuprates. In
consistency with this previous interpretation, our present Nernst
effect studies suggest that two-types of carriers exist in PCCO
for all cerium dopings we studied. Quantitative analysis of how
two types of carriers are introduced in the system, and the
variation of their concentration with cerium and oxygen doping
requires further systematic studies. We should also mention that
the Nernst effect of some hole-doped compounds (especially at
optimal doping) has been measured to high accuracy in the normal
state up to room temperature. These experiments have shown that
the Nernst signal decreases dramatically just above T$_c$, and
remains less than 50 nV/K for temperatures up to 330 K
~\cite{clayhold2,gasumyants}. These signal levels are much smaller
than what is found in PCCO suggesting one-type of carrier in the
hole-doped cuprates.

Whether the fluctuation region observed in the under-doped PCCO is
related to the pseudogap state is an important issue. The
experiments that studied the pseudogap state in electron-doped
compounds have not yet produced conclusive results about either
the magnitude or the onset temperature (T*) of the pseudogap. The
experiments in which evidence for the pseudogap state has been
claimed are tunnelling spectroscopy(T* $\leq$
T$_c$)~\cite{biswas,allf}, optical conductivity (T* $>$ 292
K~\cite{singley} to T*=110 K~\cite{homes}),
photoemission~\cite{armitage2}, and Raman
spectroscopy~\cite{koitzsch} (T*=220 K). Unlike the experiments on
hole-doped compounds that showed the pseudogap to be near the
($\pi$, 0) region, the location of the pseudogap on the Fermi
surface is also controversial for the electron-doped cuprates.
Photoemission showed gap-like features near the intersection of
the underlying Fermi surface with the antiferromagnetic Brillouin
zone boundary whereas Raman spectroscopy showed a suppression of
spectral weight for the B$_{2g}$ Raman response in the vicinity of
($\pm \pi/2, \pm \pi/2$).

\begin{figure}
\centerline{\psfig{figure=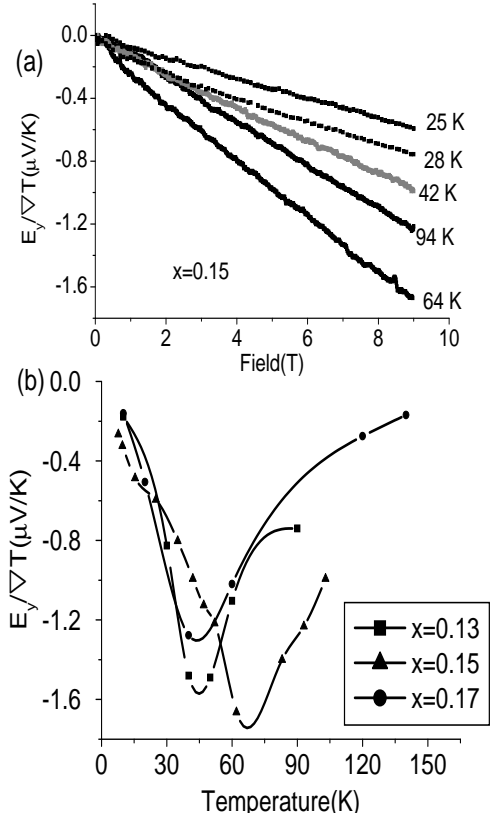,width=8.0cm,height=12.0cm,clip=}}
\caption{\textbf{(a)} High temperature Nernst effect as a function
of magnetic field for the optimally-doped sample. \textbf{(b)} The
temperature dependence of the Nernst effect at H=9 T for all
dopings. }
\end{figure}

Our Nernst effect data does not show a strong signal that could be
related to a pseudogap. For example, in the hole-doped compounds
where the anomalous Nernst effect has been
observed~\cite{xu,wang1,wang2,capan}, there is no distinctive
feature in the Nernst signal when crossing T$_c$ (i.e. T$_c$ does
not seem to be a special temperature). This suggests that these
excitations, which could originate at the pseudogap temperature
(T$^*$) dominate the signal around T$_c$. However, we should
mention that this type of behavior is not found in all hole-doped
cuprates. For some cuprates in which a pseudogap has been
observed, the Nernst effect does actually show a transition from a
large mixed state signal below T$_c$ to an almost zero normal
state signal just above T$_c$~\cite{clayhold,gasumyants}. This
issue will be discussed further in the summary section below. Our
Nernst effect data shows a similar behavior around T$_c$ to these
systems, i.e. the distinctive vortex Nernst signal goes to a
minimum and a clear normal state signal (linear in field) appears
just above T$_c$ (see Fig.2b for example). Therefore, we conclude
that there is no pseudogap state with associated superconducting
fluctuations in this regime of the electron-doped superconductors.
Of course a pseudogap of some other origin is possible.
Quantitatively for the under-doped sample at T$\approx$ 15K, where
an anomalous signal would be expected, the normal state
contribution is around 100 nV/K, which is small compared to the
vortex-like signal of several $\mu$V/K around T$_c$ for
under-doped La$_{2-x}$Sr$_{x}$CuO$_{4}$(LSCO)
~\cite{wang1}.However, we can not rule out the existence of a weak
pseudogap signal that is dominated by the normal state
(two-carrier) Nernst signal.

We now discuss the H$_{c2}$(T) extracted from the Nernst signal
(see Fig.4-a). The dashed lines in Fig.4-b show our method of
extracting H$_{c2}$(T). The uncertainty in the value of
H$_{c2}$(T) is found from the difference between the point of
intersection of the dashed lines and the point one would get from
extrapolating the vortex Nernst signal to zero. In our case
extrapolating the vortex Nernst signal to zero is the same as
extrapolating $S_\phi$, the transport entropy per unit length of
flux line, to zero since the flux flow resistivity is constant in
the relevant field range ($S_\phi=\phi_o e_y/\rho_{ff}$, where
$\rho_{ff}$ is the flux flow resistivity and $\phi_o$ is flux
quantum). Due to the complications of extracting the H$_{c2}$(T)
from $S_\phi$ that are detailed in Ref.~\cite{gollnik} (usually
H$_{c2}$(T) is overestimated in this method), H$_{c2}$(T) is not
extracted from $S_\phi$. In particular it was shown that
determining H$_{c2}$(T) from $S_\phi$ does not work at all for
under-doped NCCO ~\cite{gollnik}. Therefore the errors in the
value of H$_{c2}$(T) are taken large enough to take into account
this uncertainty. Considering the small difference between the
H$_{c2}$(T) values one would get by using different methods to
determine it, some of the important results of this study would be
valid in any of the methods used. One of these results is that
H$_{c2}$(0) increases with decreasing doping, since for a given
T/T$_c$ the signature of the normal state is seen at a larger
field as the doping decreases. The other conclusion that would not
change by the uncertainty in determining H$_{c2}$(T) is that the
fluctuation regime becomes narrower as the doping increases. This
can be seen by comparing the close proximity of the vortex Nernst
peak and the linear field dependent normal state contribution in
the over-doped sample vs the broad transition region between these
two typical regimes in the under-doped compound. However, one
conclusion that would change for the under-doped compound is the
linear temperature dependence of H$_{c2}$(T).  Using $S_\phi$ to
determine H$_{c2}$(T) would make it very difficult to observe any
systematic temperature dependence for H$_{c2}$(T) as was also
found in Ref.~\cite{gollnik}.

The H$_{c2}$(T) of the optimally-doped sample shows a linear
temperature dependence in the range of our Nernst effect data.
H$_{c2}$(0) is estimated using the Helfand-Werthamer
formula~\cite{werthamer}
\begin{equation}
H_{c2}(0)\approx 0.7 \times T_c \times \frac{dH_{c2}}{dT},
\end{equation}
where $\frac{dH_{c2}}{dT}$ is measured at T$_c$. H$_{c2}$(0) for
optimal doping is found to be 6.3$\pm$0.2 T, and therefore the
coherence length of the optimally-doped sample is $\xi(0)\approx
75\pm2\AA$ (from $\xi^2(0)=\frac{\phi_o}{2 \pi H_{c2}(0)}$ ) . The
H$_{c2}$(T) of the over-doped sample also shows a linear
temperature dependence except for T$>$13 K where the
superconducting-to-normal state transition starts. Using the
Helfand-Werthamer formula H$_{c2}$(0) is found to be 3.7$\pm0.4$
T, and $\xi(0)\approx 109\pm6\AA$.  Due to the broad fluctuation
region, where the Nernst signal had almost no field dependence, it
was more difficult to determine H$_{c2}$(T) for the under-doped
sample.  However, the fact that the normal state linear field
dependence of the Nernst signal in the under-doped compound is
observed at fields larger than that in the optimally-doped one
suggests that H$_{c2}$(T)  is larger in the under-doped compound.
A Helfand-Werthamer extrapolation to the H$_{c2}$(T) vs T data for
the under-doped compound yields H$_{c2}$(0)=7.1$\pm$0.5 and
$\xi(0)\approx 71\pm3\AA$. For a summary of these results see
Table 1.

\begin{figure}
\centerline{\psfig{figure=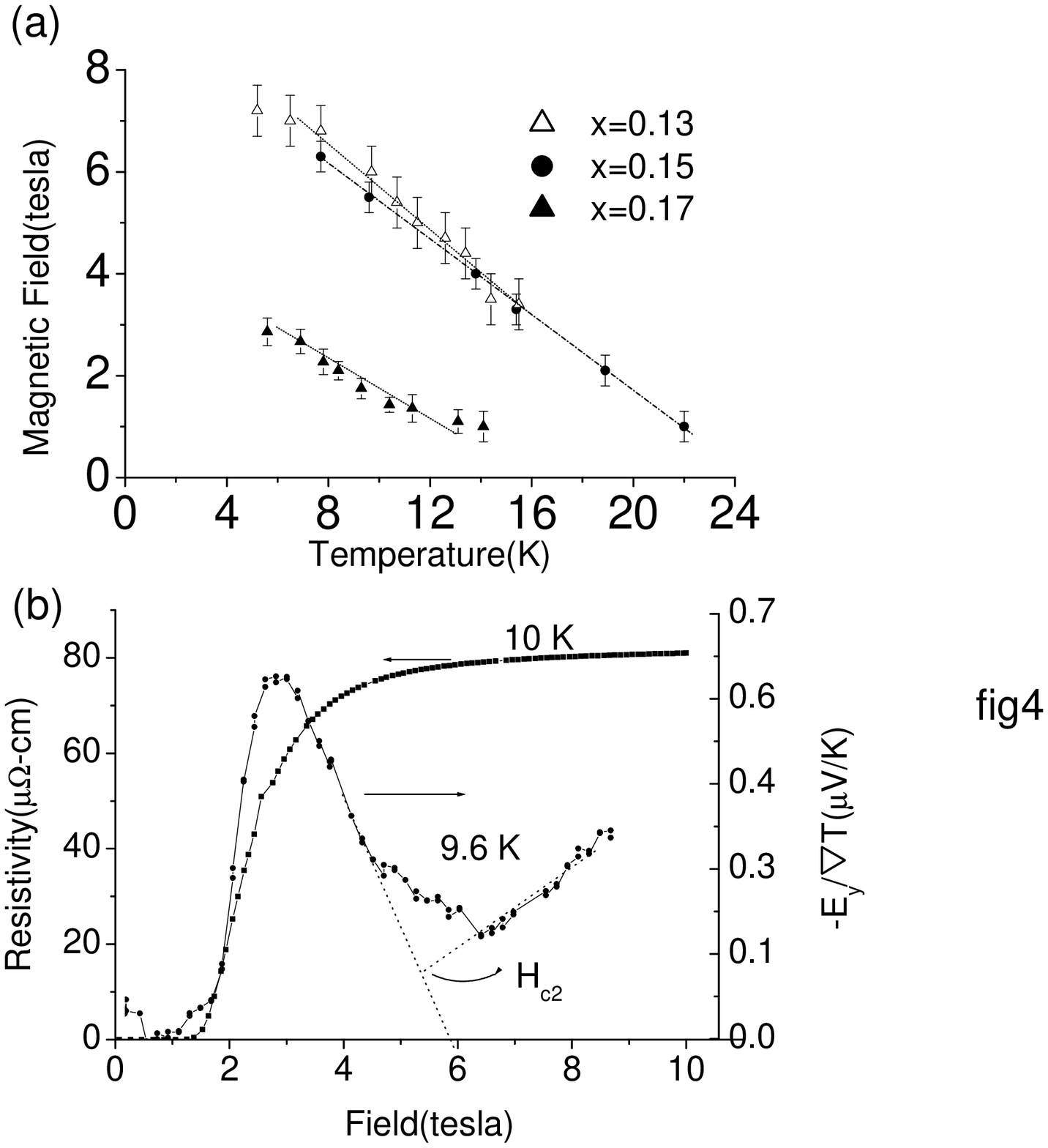,width=8.0cm,height=10.0cm,clip=}
} \caption{\textbf{(a)} The upper critical field H$_{c2}(T)$
extracted from the Nernst effect data of Fig.2 (and other data
omitted from Fig.2 for clarity). \textbf{(b)} Comparison of Nernst
effect and resistivity in terms of H$_{c2}$ for x=0.15 sample. The
dashed lines show the method used to extract H$_{c2}$.}
\end{figure}

Another important point that we should mention about the upper
critical field is the difference in the sensitivity of the Nernst
effect and resistivity in determining H$_{c2}$. Nernst effect is
very sensitive to superconducting fluctuations which are difficult
to observe in resistivity. This is particularly clear in the
under-doped compound in which the onset of the normal state
contribution is preceded by a wide fluctuation(Fig.2-a) regime in
the Nernst effect whereas the resistivity in the same field range
is basically flat (Fig.1). Resistivity measurements on PCCO and
NCCO have shown the H$_{c2}$ of the under-doped compound to be
smaller than that of the optimal-doped compound~\cite{fournier2}
(this can also be seen in the inset of Fig.1). This would imply
that the magnitude of the superconducting gap is larger in the
optimally-doped compound. However, point-contact tunnelling
experiments on similar samples have shown that the superconducting
gap amplitude is larger in the under-doped compound compared to
the optimally-doped one~\cite{biswas}. Our Nernst effect data
explains this contradiction by the insensitivity of the
resistivity to superconducting fluctuations, and implies that
resistivity is not a proper measurement for determining H$_{c2}$
in agreement with the conclusion of Ref.~\cite{fournier2}.

Resistivity and Nernst effect show similar H$_{c2}$(T)  for all
dopings if the initial deviation from the normal state resistivity
is chosen as a reference for H$_{c2}$(T)(see Fig.5b). The
under-doped compound shows a larger difference between the Nernst
effect and resistivity in terms of H$_{c2}$(T), which suggests
that the fluctuation regime is broader in the under-doped
compound. A sample curve showing the superconducting-normal state
transition from resistivity and Nernst effect is shown in Fig.5-b
for the optimally-doped sample.

There are important similarities between our Nernst effect data
and the recent Nernst effect data on hole-doped
Bi-2212(Bi$_2$Sr$_2$CaCu$_2$O$_8$) and
Bi-2201(Bi$_2$Sr$_{2-y}$La$_y$CuO$_6$)~\cite{wang3}. Similar to
our results, H$_{c2}$(0) was found to increase with decreasing
doping for both single layer and double layer Bi compounds studied
in Ref~\cite{wang3}. These observations are consistent with other
experiments showing an increasing superconducting gap ($\Delta_0
\propto v_f \sqrt{H_{c2}}$, where $v_f$ is the Fermi velocity)
amplitude with decreasing doping both for the n-doped and the
p-doped cuprates~\cite{gollnik,ino}.

\begin{table}
\centerline{
\psfig{figure=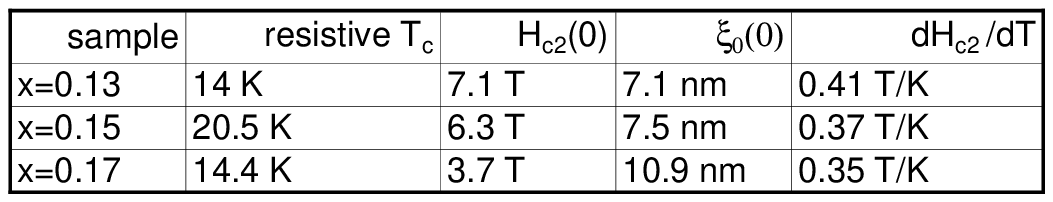,width=10.0cm,height=3.0cm,clip=} }
\caption{Summary of our results.}
\end{table}
{\bf SUMMARY}

In contrast to the hole-doped cuprates where an anomalous Nernst
signal has been observed, the vortex Nernst signal in the
electron-doped PCCO does not persist above T$_c$ or H$_{c2}$ for
over and optimal dopings. The T$_c$ and H$_{c2}$ extracted from
our Nernst effect measurements for these cerium dopings are
similar to those obtained from resistivity if the start of the
resistive superconducting transition is chosen as a reference for
T$_c$ or H$_{c2}$(T). The under-doped compound shows a broader
fluctuation regime, and therefore the superconducting to normal
state transition looks different in Nernst effect and resistivity.
Above $T_c$ the temperature dependence of the Nernst voltage is
very similar for different dopings, and the magnitude of the
Nernst signal is too large to be explained by a one-carrier model.
These results are consistent with our previous experiments on NCCO
which were interpreted as evidence for the existence of a
two-carrier transport in these materials~\cite{fournier}.

The different behavior of the Nernst effect beyond the resistive
T$_c$ (or H$_{c2}$) for n-doped and the p-doped cuprates in which
an anomalous Nernst signal is observed in the optimal and
over-doping is a puzzling problem that remains to be resolved.
However, it is clear that the large Nernst signal seen in the
normal state (T$>$T$_c$) of the n-doped cuprates has a different
origin than the anomalous Nernst signal observed in the p-doped
compounds. In our data we see a clear distinction between the
vortex Nernst effect contribution (a peak in the superconducting
state) and the normal state contribution which is linear in
magnetic field and which increases with temperature for T$>$T$_c$
up to $\sim$ 30 K above T$_c$. In contrast, the anomalous Nernst
signal observed in some of the p-doped compounds is not distinct
(there is no feature at or around T$_c$ that would distinguish the
two contributions) from the vortex Nernst contribution, and the
signal decreases with temperature for T$>$T$_c$ up to 50 K above
T$_c$~\cite{wang1}.

In conclusion, we see a possible explanation in terms of
superconducting fluctuations that can reconcile the n-doped and
p-doped Nernst experiments. Non-Gaussian fluctuations in the phase
of the superconducting order parameter are dominant between T$_c$
and the mean field critical temperature T$^{MF}_c$, but between
this T$^{MF}_c$ and the onset of the anomalous Nernst signal,
T$_{\nu}$, fluctuations both in amplitude and phase of the order
parameter should be considered in order to explain the anomalous
Nernst effect~\cite{ussishkin}. Vortex-like excitations above
T$_c$ might be an ambiguous way of describing this phenomenon
since at such conditions (high density of fluctuations) the idea
of a vortex becomes unclear. At temperatures T$>$T$^{MF}_c$
fluctuations in the amplitude of the order parameter are also
important. Hence, this would make a vortex description of such
fluctuations questionable since a certain amplitude stability is
required for a vortex to be created. These fluctuations are
smaller in the electron-doped cuprates  due to two main reasons:
\newline \textbf{1.} The effect of amplitude fluctuations is
smaller in the n-doped cuprates because of a larger coherence
length ($\sim$ 5 times larger in PCCO compared to LSCO). \newline
\textbf{2.} The phase fluctuations that dominate around T$_c$ for
the hole-doped cuprates are smaller in electron-doped compounds
since the phase stiffness temperature is comparable to the
superconducting gap amplitude in these materials. For more details
about this issue see Ref~\cite{carlson}.

Another important issue that should be reconciled with the results
of other experiments is the relation of the anomalous Nernst
signal to the pseudogap. There are two points that should be
mentioned. The first is that the onset temperature (T$_{\nu}$) of
the anomalous Nernst signal is still much less than the pseudogap
onset temperature observed in NMR or optical conductivity
experiments for all the hole-doped cuprates in which an anomalous
Nernst signal has been observed. This could imply two things:
either the pseudogap observed in NMR measurements is different and
independent of the anomalous Nernst signal or there is more than
one source for a pseudogap-like behavior (i.e. multiple
pseudogaps). The fact that the anomalous Nernst signal is more
pronounced in the under-doped regime, similar to the pseudogap
observed in other experiments, suggests that the two phenomena are
related and hence the pseudogap-like behavior is a result of more
than one mechanism. Another important fact that would support this
idea is that the anomalous Nernst signal has not been observed in
all the hole-doped cuprates that show evidence for a pseudogap
above T$_c$. These two different behavior can be reconciled by a
multiple pseudogaps model, since the absence of strong
superconducting fluctuations would not be the only way for the
creation of a pseudogap in this model. In fact there are several
proposals for the pseudogap crossover that have nothing to do with
superconductivity~\cite{varma}. But none of these proposals have
yet explained the origin of a large Nernst signal for $T_c<T<T^*$.
The second point that should be mentioned is related to the
pseudogap phenomena in the electron-doped cuprates. Tunnelling
studies on electron-doped PCCO show a pseudogap that has an onset
temperature T*$<$T$_c$~\cite{biswas,allf}. It is not clear at this
moment if this behavior can be reconciled with the other
experiments that suggest a pseudogap at temperatures much higher
than T$_c$. Again considering multiple origins for the pseudogap
could explain these different experiments which probe different
physical properties. However, having a T*$<$T$_c$ would be another
plausible explanation for why an anomalous Nernst signal is not
observed in electron-doped cuprates above T$_c$.

Clearly, more work on the nature of the pseudogap state in the
n-doped cuprates needs to be done before any conclusive
explanation of the n-doped and p-doped Nernst effect data can be
made. At the present time a superconducting fluctuation induced
anomalous Nernst effect would appear to be most consistent with
all the known experimental data.

\textbf{ACKNOWLEDGEMENTS}\newline We would like to thank Steven A.
Kivelson and Iddo Ussishkin for many helpful discussions. This
work was supported by the NSF DMR 01-02350.

\end{document}